\documentclass[12pt,draftcls,onecolumn,a4paper]{IEEEtran}
\usepackage{amsmath,epsfig,amssymb}

\newtheorem{thm}{Theorem}
\newtheorem{lemma}{Lemma}

\long\def\comment#1{}

\newfont{\bb}{msbm10 scaled 1000}
\newcommand{\CC}{\mbox{\bb C}}

\newcommand{\EE}{\mbox{\bb E}}

\newfont{\bbsmall}{msbm10 scaled 700}

\newcommand{\nv}{{\bf n}}

\newcommand{\uv}{{\bf u}}

\newcommand{\xv}{{\bf x}}
\newcommand{\yv}{{\bf y}}

\newcommand{\Am}{{\bf A}}
\newcommand{\Bm}{{\bf B}}

\newcommand{\Dm}{{\bf D}}
\newcommand{\Em}{{\bf E}}
\newcommand{\Fm}{{\bf F}}
\newcommand{\Gm}{{\bf G}}
\newcommand{\Hm}{{\bf H}}

\newcommand{\Nm}{{\bf N}}

\newcommand{\Qm}{{\bf Q}}

\newcommand{\Sm}{{\bf S}}

\newcommand{\Wm}{{\bf W}}
\newcommand{\Vm}{{\bf V}}
\newcommand{\Xm}{{\bf X}}
\newcommand{\Ym}{{\bf Y}}
\newcommand{\Zm}{{\bf Z}}

\newcommand{\Hhm}{{\bf \widehat{ H}}}

\newcommand{\Vhm}{{\bf \widehat{ V}}}

\newcommand{\Cc}{{\cal C}}

\newcommand{\Gc}{{\cal G}}

\newcommand{\trace}{{\hbox{tr}}}

\renewcommand{\arg}{{\hbox{arg}}}

\newcommand{\herm}{{\sf H}}

\def\ImgWidth{11cm} 

\title{Limited Feedback-based Block Diagonalization for the MIMO Broadcast Channel}
\author{\authorblockN{Niranjay Ravindran and Nihar Jindal\\}
\authorblockA{University of Minnesota\\
Minneapolis MN, 55455 USA\\
Email: {\{ravi0022, nihar\}@umn.edu}\\
}
}

\begin{document}

\maketitle
\begin{abstract}
Block diagonalization is a linear precoding technique for the multiple antenna broadcast (downlink) channel that involves transmission of multiple data streams to each receiver such that no multi-user interference is experienced at any of the receivers. This low-complexity scheme operates only a few dB away from capacity but requires very accurate channel knowledge at the transmitter. We consider a limited feedback system where each receiver knows its channel perfectly, but the transmitter is only provided with a finite number of channel feedback bits from each receiver. Using a random quantization argument, we quantify the throughput loss due to imperfect channel knowledge as a function of the feedback level. The quality of channel knowledge must improve proportional to the SNR in order to prevent interference-limitations, and we show that scaling the number of feedback bits linearly with the system SNR is sufficient to maintain a bounded rate loss. Finally, we compare our quantization strategy to an analog feedback scheme and show the superiority of quantized feedback.
\end{abstract}


\section{Introduction}
\label{sec:intro} 

In multiple antenna broadcast (downlink) channels, transmit antenna arrays can be used to simultaneously transmit data
streams to receivers and thereby significantly increase throughput. Dirty paper coding (DPC) is capacity
achieving for the MIMO broadcast channel \cite{WEIN04}, but this technique has a very high level of complexity. Zero
Forcing (ZF) and Block Diagonalization (BD) \cite{CHOI04} \cite{SPEN04} are alternative low-complexity transmission
techniques.  Although not optimal, these linear precoding techniques utilize all available spatial degrees of freedom and perform measurably close to DPC in many scenarios \cite{LEE06}.

If the transmitter is equipped with $M$ antennas and there are at least $M$ aggregate receive antennas, zero-forcing involves transmission of $M$ spatial beams such that independent, de-coupled data channels are created from the transmit antenna array to $M$ receive antennas distributed amongst a number of receivers.  Block diagonalization similarly involves transmission of $M$ spatial beams, but the beams are selected such that the signals received at different receivers, but not necessarily at the different antenna elements of a particular receiver, are de-coupled.  For example, if there are $M/2$ receivers with two antennas each, then two beams are aimed at each of the receivers.  If ZF is used, an independent and de-coupled data stream is received on each of the $M$ antennas. If BD is used, the streams for different receivers do not interfere, but the two streams intended for a single receiver are generally not
aligned with its two antennas and thus post-multiplication by a rotation matrix (to align the streams) is generally required before decoding.

In order to correctly aim the transmit beams, both schemes require perfect Channel State Information at the Transmitter (CSIT). Imperfect CSIT leads to incorrect beam selection and therefore multiuser interference, which ultimately leads to a throughput loss. Unlike point to point MIMO systems where imperfect CSIT causes only an SNR offset in the capacity vs.\@ SNR curve, the level of CSIT affects the slope of the curve and hence the \textit{multiplexing} gain in broadcast MIMO systems. We consider the case when the CSI is known perfectly at the receiver and is communicated to the transmitter through a limited feedback channel and quantify the maximum rate loss due to limited feedback with BD. 

MISO systems and ZF with limited feedback are analyzed in \cite{JIND05}. Similar to the results in \cite{JIND05}, we show that scaling the number of feedback bits approximately linearly with the system SNR is sufficient to maintain the slope of the capacity vs.\@ SNR curve and hence a constant gap from the capacity of BD with perfect CSIT. The scaling factor for BD offers an advantage over ZF in terms of the number of bits required to achieve the same sum capacity. 

Rather than quantizing the CSIT into a finite number of bits and feeding this information back, the channel coefficients can also be explicitly transmitted over the feedback link. We compare this scheme to quantized feedback for an AWGN feedback channel, and show the superiority of quantized feedback.

\section{System Model}
\label{sec:sysmodel}

We consider a MIMO broadcast (downlink) system with a single transmitter or base station and $K$ receivers or users. Each user has $N$ antennas and the transmitter has $M$ antennas. The broadcast channel is described as:
\begin{equation}
\yv_k = \Hm_k^\herm\xv + \nv_k,\quad k = 1, \dots, K
\end{equation}
where $\Hm_k \in \CC^{M \times N}$ is the channel matrix from the transmitter to the $k^\text{th}$ user ($1 \leq k \leq K$) and the vector $\xv \in \CC^{M \times 1}$ is the transmitted signal. $\nv_k \in \CC^{N \times 1}$ are independent complex Gaussian noise vectors of unit variance and $\yv_k \in \CC^{N \times 1}$ is the received signal vector at the $k^\text{th}$ user. We assume  a transmit power constraint so that $E[||\xv||^2] \leq P$ $(P > 0)$. We also assume that $K = \frac{M}{N}$ (with $K \geq 2$), which implies that the aggregate number of receive antennas equals the number of transmit antennas; as a result it is not necessary to select a subset of users for transmission.

The entries of $\Hm_k$ are assumed to be i.i.d.\@ unit variance complex Gaussian random variables, and the channel is assumed to be block fading with independent fading from block to block. Each of the users are assumed to have perfect and instantaneous knowledge of their own channel matrix. The channel matrix is quantized by each user and fed back to the transmitter (which has no other knowledge of the instantaneous CSI) over a zero delay, error free, limited feedback channel. 
 
It is assumed that a uniform power allocation policy is adopted (i.e.\@, we do not perform waterfilling across streams), which is known to be asymptotically optimal for large SNR. Hence, in order to perform Block Diagonalization, it is only necessary to know the spatial direction of each user's channel, i.e.\@, the subspace spanned by the columns of $\Hm_k$, and the feedback only needs to convey this information. 

The quantization codebook used by each user is fixed beforehand and is known to the transmitter. A quantization codebook $\Cc$ consists of $2^B$ matrices in $\CC^{M \times N}$ i.e.\@ $(\Wm_1, \dots, \Wm_{2^B})$, where $B$ is the number of feedback bits allocated per user. The quantization of a channel matrix $\Hm_k$, say $\Hhm_k$, is chosen from the codebook $\Cc$ according to the following rule:
\begin{equation} \label{quantproc}
\Hhm_k = \mathop{\arg \min}\limits_{\Wm\ \in\ \Cc}\ d^2\left( \Hm_k, \Wm \right)
\end{equation}
where $d\left( \Hm_k, \Wm \right)$ is the distance metric. Here, we consider the \textit{chordal distance} \cite{CONW96}:
\begin{equation}
d\left( \Hm_k, \Wm \right) = \sqrt{\sum\limits_{j=1}^N \sin^2\theta_j}
\end{equation}
where the $\theta_j$'s are the principal angles between the two subspaces spanned by the columns of the matrices $\Hm_k$ and $\Wm$ \cite{CONW96}. As the principal angles depend only on the subspaces spanned by the columns of the matrices, it can be assumed that the elements of $\Cc$ are unitary matrices (i.e.\@ $\Wm^\herm\Wm = {\bf I}_N\ \forall\ \Wm \in \Cc$), without loss of generality. An alternate form for the chordal distance is $d^2\left( \Hm_k, \Wm \right) = N - \trace\left( \widetilde{\Hm}_k^\herm\Wm\Wm^\herm\widetilde{\Hm}_k \right)$, where $\widetilde{\Hm}_k$ forms an orthonormal basis for the subspace spanned by $\Hm_k$. Note that other distance metrics may also be considered, but we do not investigate this further in this work. No channel magnitude information is fed back to the transmitter.

\section{Background}
\label{sec:background}

\subsection{Block Diagonalization}
\label{ssec:bdiag}

The Block Diagonalization strategy, when perfect CSI is available at the transmitter, involves linear precoding that suppresses the interference at each user due to all other users (but does not suppress interference due to different antennas for the same user). If $\uv_k \in \CC^{N \times 1}$ contains the $N$ complex (data) symbols intended for the $k^\text{th}$\ ($1 \leq k \leq K$) user and $\Vm_k \in \CC^{M \times N}$ is the precoding matrix, then the transmitted vector is given by:
\begin{equation}
\xv = \sum\limits_{k = 1}^K \Vm_k\uv_k
\end{equation}
and the received signal at the $k^\text{th}$ user is given by:
\begin{equation}
\yv_k = \Hm_k^\herm \Vm_k\uv_k + \sum\limits_{j = 1, j \neq k}^K \Hm_k^\herm \Vm_j\uv_j + \nv_k
\end{equation}

The $\sum\limits_{j = 1, j \neq k}^K \Hm_k^\herm \Vm_j\uv_j$ term represents the multi-user interference at user $k$. In order to maintain the power constraint, it is assumed that $\Vm_k^\herm\Vm_k = {\bf I}_N$ and $E\left[||\uv_k||^2\right] \leq \frac{P}{M}$, for $k = 1, \dots, K$.

Following the BD procedure, each $\Vm_k$ is chosen such that $\Hm_j^\herm\Vm_k$ is ${\bf 0},\ \forall k \neq j$. This amounts to determining an orthonormal basis for the left null space of the matrix formed by stacking all $\{\Hm_j\}_{j \neq k}$ matrices together. This reduces the interference terms in equation (\ref{eqn:rxbd}) to zero at each user. This is different from Zero Forcing where each complex symbol to be transmitted to the $m^\text{th}$ antenna (among the $N$ antennas, i.e.\@, $m = 1, \dots, N$) of the $k^\text{th}$ user is precoded by a vector that is orthogonal to all the columns of $\Hm_j, j \neq k$, as well as orthogonal to all but the $m^\text{th}$ column of $\Hm_k$.

However, zero interference can only be achieved with perfect knowledge of $\{\Hm_k\}_{k=1}^K$ at the transmitter. In the case of limited feedback, when only a quantized version of the subspace spanned by the columns of each $\Hm_k$ is available at the transmitter, namely $\Hhm_k$, we use a naive strategy where the precoding matrices are selected by treating $\Hhm_1, \dots, \Hhm_K$ as the true channels while performing BD. To distinguish these precoding matrices from those selected with perfect CSIT, we denote these matrices as $\Vhm_1, \dots, \Vhm_K$, where each $\Vhm_k$ is chosen such that $\Hhm_j^\herm\Vhm_k = {\bf 0}\ \forall k \neq j$. Thus, $\Hm_j^H\Vhm_k \neq 0$ in general, which leads to residual interference terms and a loss in throughput. The received signal in the case of limited feedback is thus written as:
\begin{equation}\label{eqn:rxbd}
\yv_k = \Hm_k^\herm \Vhm_k\uv_k + \sum\limits_{j = 1, j \neq k}^K \Hm_k^\herm \Vhm_j\uv_j + \nv_k
\end{equation}

\subsection{Random Quantization Codebooks}
\label{ssec:randcode}

Since the design of optimal quantization codebooks for the given
distance metric is a very difficult problem, we instead study
performance averaged over \textit{random} quantization codebooks.
The Grassmann manifold is the set of all $N$ dimensional
subspaces (or planes) passing through the origin, in an $M$ dimensional space. This is denoted by
$\Gc_{M, N}$. We consider complex Euclidean subspaces in this work. Each of the $2^B$ unitary matrices making up the random quantization codebook are chosen independently and are uniformly distributed over $\Gc_{M, N}$ \cite{DAI06} \cite{marzetta1999cmm}. We alternatively refer to this uniform distribution as the isotropic distribution in the respective space. A random element drawn from this distribution (over $\Gc_{M, N}$) can be generated by generating an $M \times N$ matrix with i.i.d.\@ complex Gaussian elements and then forming a specific orthonormal basis for the $N$ dimensional subspace spanned by the matrix (e.g.\@, through a QR decomposition).

We analyze the performance  averaged over all possible random codebooks. The distortion or error associated with a given codebook $\Cc$ for the quantization of $\Hm_k \in \CC^{M \times N}$ is defined as:
\begin{equation}
D \mathop{=}\limits^\Delta \EE \left[ d^2(\Hm_k,\Hhm_k)  \right] = \EE \left[ \mathop{\min\limits_{\Wm \in \Cc}}\ d^2(\Hm_k,\Wm) \right],
\end{equation}
where $\Hhm_k$ is the quantization of $\Hm_k$. It is shown in \cite{DAI06} that $D \leq \overline{D}$ where,
\begin{equation} \label{eqn:D}
\overline{D} = \frac{\Gamma(\frac{1}{T})}{T} (C_{MN})^{-\frac{1}{T}} 2^{-\frac{B}{T}} + N \exp\left[ -(2^BC_{MN})^{1-a} \right] ,
\end{equation}
for a codebook of size $2^B$. Here, $T = N (M - N)$ and $a \in (0, 1)$ is a real number between $0$ and $1$ chosen such that $\left(  C_{MN}2^B  \right)^{-\frac{a}{T}} \leq 1$. $C_{MN}$ is given by $\frac{1}{T!}\ \prod\limits_{i = 1}^N\ \frac{(M - i)!}{(N - i)!}$. The second (exponential) term in (\ref{eqn:D}) can be neglected for large $B$. For systems where $N = 2$ or $3$, the exponential term may be neglected for most practical cases.

\section{Analysis and Results}
\label{sec:tanal}

In this section, we analyze the achievable throughput of the limited feedback-based system described so far. We first describe some preliminary mathematical results.

\subsection{Preliminary Calculations}
\label{ssec:prelim}

\begin{lemma} \label{lemma1}
The quantization $\Hhm_k$ of the channel $\Hm_k$ admits the following decomposition:
\begin{equation} \label{lem1}
\widetilde{\Hm}_k = \Hhm_k \Xm_k \Ym_k + \Sm_k \Zm_k
\end{equation}
where
\begin{itemize}
\item $\widetilde{\Hm}_k \in \CC^{M \times N}$ is an orthonormal basis for the subspace spanned by the columns of $\Hm_k$,
\item $\Xm_k \in \CC^{N \times N}$ is unitary and distributed uniformly over $\Gc_{N, N}$,
\item $\Zm_k \in \CC^{N \times N}$ is upper triangular with positive diagonal elements, satisfying $\trace(\Zm_k^\herm\Zm_k) = d^2\left( \Hm_k, \Hhm_k \right)$,
\item $\Ym_k \in \CC^{N \times N}$ is upper triangular with positive diagonal elements and satisfies $\Ym_k^\herm\Ym_k = {\bf I}_N - \Zm_k^\herm\Zm_k$, and
\item $\Sm_k \in \CC^{M \times N}$ is an orthonormal basis for an isotropically distributed (complex) $N$ dimensional plane in the $M-N$ dimensional left nullspace of $\Hhm_k$.
\end{itemize}

Moreover, the quantities $\Ym_k$, $\Hhm_k$ and $\Xm_k$ are distributed independent of each other, as are the pair $\Sm_k$ and $\Zm_k$. This decomposition is a generalization of the decomposition in \cite{JIND05}, which was for the specific case of $N = 1$. Similar to \cite{JIND05}, the matrix $\Zm_k$ represents the quantization error.

\end{lemma}

\vspace{12pt}
\begin{proof} See Appendix \ref{lem1proof}. \end{proof}
\vspace{12pt}

A direct application of Lemma \ref{lemma1} allows us to bound the rate loss due to limited feedback. This decomposition also allows us to perform low complexity Monte-Carlo simulations for evaluating the performance of random quantization codebooks, even for very large $B$, as described in detail in Section \ref{sec:simmethod}.

\subsection{Throughput analysis for quantized feedback}
\label{ssec:ffqual}

In the case of perfect CSIT and BD, the transmitter has the ability to suppress all interference terms giving a {\emph{per user}} ergodic rate of:
\begin{equation} \label{eqn:R_CSIT}
R_\textsc{CSIT-BD}(P) = \EE \left[ \ \log_2 \left| {\bf I}_N + \frac{P}{M}\Hm_k^\herm\Vm_k\Vm_k^\herm\Hm_k \right|\ \right]
\end{equation}
where $k$ is any user from $1, \dots, K$. The expectation is carried out over the distribution of $\Hm_k$.

For limited feedback of $B$ bits per user, multiuser interference cannot be completely canceled and this leads to residual interference power.  The per-user rate (throughput) is given by:
\begin{eqnarray}
R_\textsc{Quant}(P) & = & \EE\left[ I(\uv_k; \yv_k \vert \Hm_k) \right]\\
& = & \EE \left[ \log_2\left| {\bf I}_N + \frac{P}{M} \left( {\bf I}_N + \frac{P}{M} \sum\limits_{j = 1, j \neq k}^K \Hm_k^\herm\Vhm_j\Vhm_j^\herm\Hm_k \right)^{-1} \Hm_k^\herm\Vhm_k\Vhm_k^\herm\Hm_k \right| \right]\\
& = & \EE \left[ \log_2 \left| {\bf I}_N + \frac{P}{M} \sum\limits_{j = 1}^K \Hm_k^\herm\Vhm_j\Vhm_j^\herm\Hm_k \right| \right] - \nonumber\\ &  & \EE \left[ \log_2 \left| {\bf I}_N + \frac{P}{M} \sum\limits_{j = 1, j \neq k}^K \Hm_k^\herm\Vhm_j\Vhm_j^\herm\Hm_k \right| \right] \label{eqn:R_FB}
\end{eqnarray}
where $k$ is any user between $1$ and $K$ and the expectation is carried out over the channel distribution as well as random codebooks $\Cc$.

\begin{thm} \label{thm:1}
The rate loss incurred per user due to limited feedback with respect to perfect CSIT using Block Diagonalization can be bounded from above by:
\begin{eqnarray*}
\Delta R_\textsc{Quant}(P) & = & \left[  R_\textsc{CSIT-BD}(P) - R_\textsc{Quant}(P) \right] \\
& \leq & N\ \log_2 \left(1 + \frac{P}{N} D \right)
\end{eqnarray*}
\end{thm}

\vspace{12pt}
\begin{proof} See Appendix \ref{thm1proof}. \end{proof}
\vspace{12pt}

This provides a bound on the rate loss per user\footnote{Note that a factor of $N$ was erroneously omitted from this bound when this result was stated in \cite{ravindran2007mbc}.}. Furthermore, $D$ can be upper bounded tightly by $\overline{D}$ from (\ref{eqn:D}).

\subsection{Controlling feedback quality}
\label{ssec:incfqual}

\begin{figure}[htb]
\begin{center}
\includegraphics[width=\ImgWidth]{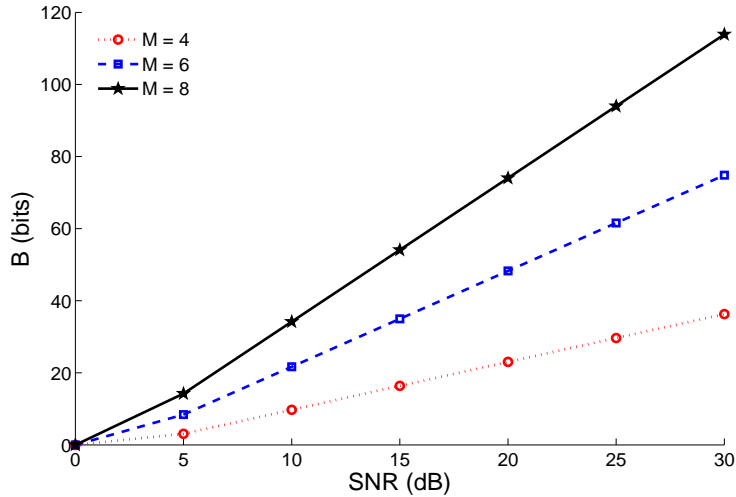}
\caption{Sufficient number of bits for a gap of 3 dB relative to BD with perfect CSIT, for $N = 2$ and $M = 4, 6$ and $8$}\label{fig0}
\end{center}
\end{figure}

If $B$ is kept fixed and the SNR is taken to $\infty$, it is easy to see that residual interference will eventually overwhelm signal power, and this leads to a bounded throughput (i.e.\@, zero multiplexing gain). Therefore, it is of interest to determine how fast $B$ must grow with SNR in order to prevent this behavior and to maintain a bounded rate loss relative to a perfect CSIT system.

\begin{thm} \label{thm:2}
In order to bound the per-user rate loss $\Delta R_\textsc{Quant}(P)$ from above by $\log_2(b) > 0$, it is sufficient for the number of feedback bits per user to be scaled with SNR as:
\begin{eqnarray} \label{eqn:B}
B \approx & \frac{N (M - N)}{3}P_{dB} - N (M - N) \log_2(N(b^{\frac{1}{N}}-1))\ + \nonumber\\
& N (M - N) \log_2 \left[ \frac{\Gamma(\frac{1}{N(M-N)})}{N(M-N)} \right] - \log_2(C_{MN})
\end{eqnarray}
\end{thm}

\begin{proof}
This expression can be found by equating the upper bound from Theorem \ref{thm:1} with $\log_2 b$ and solving for $B$ as a function of $P$. Solving this numerically will yield the number of bits sufficient for a maximum rate loss of $\log_2 b$. We assume that $B$ is large enough to neglect the exponential term in the expression for $\overline{D}$ from (\ref{eqn:D}), which yields the above approximation.
\end{proof}

The total contribution of the term containing the logarithm of the gamma function is very small and can usually be neglected. To maintain a system throughput loss of $M$ bps/Hz, which corresponds to an SNR gap of no
more than $3$ dB with respect to BD with perfect CSIT, it is sufficient to scale the bits as:
\begin{equation} \label{eqn:BDSca}
B \approx \frac{N (M - N)}{3}P_{dB} - \log_2(C'_{MN})
\end{equation}
where $C'_{MN} = N^{N(M-N)} C_{MN}$. Figure \ref{fig0} shows the sufficient number of bits required to maintain this level of performance, when $N = 2$ and $M = 4, 6$ and $8$.

The pre-log factor (i.e.\@ the factor that multiplies the SNR in dB) is $N (M-N)$ rather than $MN$, which is intuitively because the space of $N$ dimensional subspaces in an $M$ dimensional space has a dimensionality of $N( M-N)$

\section{Performance Comparison and Numerical Results}
\label{sec:perfcomp}

\subsection{Zero forcing vs.\@ Block diagonalization}
\label{ssec:ZF}

Zero forcing is simple low-complexity linear precoding strategy, and it is important to compare the performance of these two schemes under the
presence of limited feedback.  Zero forcing for a MIMO broadcast system with $K$ users and $N$ antennas per user is equivalent to a $KN = M$ user system with a single antenna per user. The feedback scaling law for such a system is derived in \cite{JIND05} to be:
\begin{equation} \label{eqn:ZFSca}
B_{ZF} \approx \frac{(M - 1)}{3}P_{dB}
\end{equation}
to maintain an SNR gap of no more than $3$ dB with respect to ZF under perfect CSIT conditions. In this system, each user with $N$ antennas quantizes the direction of the channel vector (i.e.\@ the channel vector normalized to have norm unity) of each of the $N$ antennas separately, and feeds this back to the transmitter.

In general, if BD with perfect CSIT achieves a sum rate of $R_{CSIT-BD}(P)$ with $M$, $N$ antennas at the transmitter and each of the $\frac{M}{N}$ users respectively, and ZF achieves $R_{CSIT-ZF}(P)$ for the same system, $R_{CSIT-BD}(P)$ will eventually dominate $R_{CSIT-ZF}(P)$ by a constant amount. Thus, we see an immediate advantage of BD with respect to ZF from (\ref{eqn:BDSca}), where the pre-log factor for BD is $N(M-N)$ for $N$ antennas, or $M-N$ per user antenna. This is compared to the factor $M-1$ in (\ref{eqn:ZFSca}), which is for a lower target rate. This difference between $M-1$ and $M-N$ is perhaps due to the fact that the space of $N$ dimensional subspaces in an $M$ dimensional space has a dimensionality of $N(M-N)$ while the space of $N$ one-dimensional subspaces in an $M$ dimensional space has dimensionality $N(M-1)$. 

The rate gap between BD and ZF with perfect CSIT is given by \cite{LEE06}: 
\begin{equation}
R_g(P) = K\log_2(e) \sum_{j=1}^N\frac{N-j}{j}
\end{equation}
at high SNR. For fair comparison of the number of bits required for BD and ZF under imperfect CSIT and limited feedback, it is necessary to fix a common target rate. By setting $b = 2^{R_g(P) + R}$ in (\ref{eqn:B}) where $R_g(P)$ is the (per-user) rate gap between BD and ZF (with perfect CSIT) and $R$ the target (per-user) rate loss for the ZF system, we can compare the \textit{sufficient} number of bits required to achieve the same sum rate for both strategies. For example, $R = 1$ for a $3$ dB target offset in SNR, relative to rate achievable with ZF and perfect CSIT. This suggests a bit savings of 48\% for an $M = 6, N = 2$ system at 15 dB, and 63\% for an $M = 9, N = 3$ system with BD. The scaling law in Theorem \ref{thm:2} is slightly conservative for large $b$, and the advantage of BD is somewhat underestimated. Numerical results show that the bit savings possible with BD are even higher.

An alternative antenna combining method (when the users have multiple antennas) is proposed in \cite{jindal2007ant}, where each user receives only a single stream of data (as opposed to $N$ streams of data with BD), but uses the extra antennas to obtain a very accurate quantization of the effective channel. This effectively allows for a reduction in feedback load, and produces the same pre-log factor as BD, i.e.\@, $N(M-N)$, but needs $N$ times the number of users in the system (i.e.\@ $K = M$ where each user as $N$ antennas, rather than the $K = \frac{M}{N}$ for BD). Table \ref{tab1} compares the {\emph{sufficient}} number of bits required to achieve the same target rate, i.e.\@, 3 dB (in SNR) away from ZF with perfect CSIT, when using BD, ZF and Antenna combining for an $M = 6, N = 2$ system. ZF and BD have $K = 3$, while antenna combining has $K = 6$.

\begin{table*}[ht]
	\centering
\begin{tabular}{l|l|l|l}
\multicolumn{1}{c|}{SNR} & \multicolumn{1}{c|}{Block Diagonalization} & \multicolumn{1}{c|}{Zero Forcing} & \multicolumn{1}{c}{Antenna Combining} \\ 
\hline
\multicolumn{1}{c|}{5 dB} & \multicolumn{1}{c|}{1} & \multicolumn{1}{c|}{9} & \multicolumn{1}{c}{8} \\ 
\hline
\multicolumn{1}{c|}{10 dB} & \multicolumn{1}{c|}{7} & \multicolumn{1}{c|}{17} & \multicolumn{1}{c}{15} \\ 
\hline
\multicolumn{1}{c|}{15 dB} & \multicolumn{1}{c|}{13} & \multicolumn{1}{c|}{25} & \multicolumn{1}{c}{21} \\ 
\hline
\multicolumn{1}{c|}{20 dB} & \multicolumn{1}{c|}{20} & \multicolumn{1}{c|}{34} & \multicolumn{1}{c}{28} \\ 
\hline
\multicolumn{1}{c|}{25 dB} & \multicolumn{1}{c|}{26} & \multicolumn{1}{c|}{42} & \multicolumn{1}{c}{35} \\ 
\hline
\multicolumn{1}{c|}{30 dB} & \multicolumn{1}{c|}{33} & \multicolumn{1}{c|}{50} & \multicolumn{1}{c}{41} \\ 
\end{tabular}
	\caption{Feedback requirement (bits) for different multiple user-antenna strategies ($M = 6, N = 2$)} \label{tab1}
\vspace{-24pt}
\end{table*}

\subsection{Analog Feedback}
\label{ssec:analogfb}

We consider here the case when each user $k$ feeds back its channel $\Hm_k$ by explicitly transmitting the $MN$ complex coefficients $\left(\Hm_k\right)_{mn}, m = 1, \dots M, n = 1, \dots, N$ over the feedback channel. We assume that the uplink feedback channel is unfaded AWGN with the same SNR as the downlink (i.e.\@, $P$). Each user may transmit each coefficient effectively `$\beta$' times on the uplink, resulting in the following matrix being received at the transmitter:
\begin{eqnarray}
\Gm_k &=& \sqrt{\beta P} \Hm_k + \Nm_k .
\end{eqnarray}
Here, $\Nm_k$ represents the feedback (additive white Gaussian) noise, whose entries are independent and complex Gaussian with unit variance. As the coefficients of $\Hm_k$ are also independent and complex Gaussian with unit variance, the optimal estimator is the MMSE estimator:
\begin{eqnarray}
\breve{\Hm}_k &=& \frac{\sqrt{\beta P}}{1 + \beta P} \Gm_k, 
\end{eqnarray}
where $\breve{\Hm}_k$ is the estimate of $\Hm_k$ formed at the transmitter. It is convenient to express $\Hm_k$ in terms of the estimate $\breve{\Hm}_k$ and estimation noise as follows:
\begin{eqnarray}
\Hm_k &=& \breve{\Hm}_k + \frac{1}{\sqrt{1 + \beta P}} \Fm_k,
\end{eqnarray}
where the entries of $\Fm_k$ are also independent and complex Gaussian with unit variance, and independent of the estimator.

The beamformers $\{\breve{\Vm}_k\}_{k=1}^K$ are selected by treating $\{\breve{\Hm}_k\}_{k=1}^K$ as the `true' set of channels, and following the BD procedure. Note that the marginal distribution of the beamformers are the same as in the quantized feedback case, as the addition of independent white Gaussian noise does not affect the isotropic property. As in the case for quantized (digital) feedback, we compute the quantity:
\begin{eqnarray} \label{int_ana}
\Hm_k^\herm\breve{\Vm}_j &=& \frac{1}{\sqrt{1 + \beta P}} \Fm_k^\herm\breve{\Vm}_j
\end{eqnarray}
for $k \neq j$, which follows from the fact that $\breve{\Hm}_k^\herm\breve{\Vm}_j = {\bf 0}$ for $k \neq j$. Similar to (\ref{eqn:R_FB}), we write the rate with `analog' feedback as follows:
\begin{eqnarray}
R_\textsc{Analog}(P) = \EE \left[ \log_2 \left| {\bf I}_N + \frac{P}{M} \sum\limits_{j = 1}^K \Hm_k^\herm\breve{\Vm}_j\breve{\Vm}_j^\herm\Hm_k \right| \right] - \EE \left[ \log_2 \left| {\bf I}_N + \frac{P}{M} \sum\limits_{j = 1, j \neq k}^K \Hm_k^\herm\breve{\Vm}_j\breve{\Vm}_j^\herm\Hm_k \right| \right]
\end{eqnarray}

Similar to the proof of Theorem \ref{thm:1} and using techniques similar to those in \cite{FastCSI}, we compute a bound on the rate gap relative to BD with perfect CSIT to be:
\begin{eqnarray}
\Delta R_\textsc{Analog}(P) & = & \left[  R_\textsc{CSIT-BD}(P) - R_\textsc{Analog}(P)  \right]\\
& \leq & N\ \log_2 \left(1 + \frac{M-N}{M}  \frac{P}{1 + \beta P} \right) \label{analogbound}\\
& < & N\ \log_2 \left(1 + \frac{M-N}{M}  \frac{1}{\beta} \right) \label{analogbound2}
\end{eqnarray}

\noindent
The proof (\ref{analogbound}) bound is given in Appendix \ref{analogproof}. (\ref{analogbound2}) is obtained by letting $P \rightarrow \infty$ in (\ref{analogbound}).

In order to compare analog and quantized feedback, we measure the feedback quantity in terms of `feedback symbols' rather than bits. Although analog feedback involves effectively $\beta MN$ channel uses per user (assuming that the users have orthogonal feedback channels), it also conveys more information that the quantized case, specifically information regarding the eigenvalues and eigenvector structure, which the `subspace' information does not capture. 

Hence, for fair comparison, we equate the $\beta MN$ analog channel uses to $\beta N(M-N)$ channel symbols in the quantized case (the `subspace' information may be specified by $N(M-N)$ complex numbers). Under the simplifying assumption that error-free communication at capacity is possible, we set $B = \beta N(M-N) \log_2(1 + P)$ for $\beta N(M-N)$ channel uses of the AWGN feedback channel with SNR $P$. From Theorem \ref{thm:1}, we have:
\begin{eqnarray}
\Delta R_\textsc{Quant}(P) & \leq & N \log_2\left(1 + \frac{P}{N} \frac{\Gamma\left((N(M-N))^{-1}\right)}{N(M-N)}C_{MN}^{\left(N(M-N)\right)^{-1}} 2^{-\frac{B}{N(M-N)}}\right)\\
& = & N \log_2\left(1 + \frac{P}{(1+P)^\beta} C''_{MN}\right)
\end{eqnarray}
where $D$ has been bounded from (\ref{eqn:D}) (neglecting the exponential term), and
\begin{equation}
C''_{MN} = \frac{\Gamma\left((N(M-N))^{-1}\right)}{N^2(M-N)} C_{MN}^{\left(N(M-N)\right)^{-1}} .
\end{equation}

Our conclusions are similar to the $N = 1$ case, which was considered in \cite{caire2007}. For $\beta \approx 1$, both bounds on the rate gap (i.e.\@ for analog and quantized feedback) behave similarly, and the gap does not vanish as $P \rightarrow \infty$. For $\beta > 1$, the rate gap bound decreases rapidly (exponentially fast) for quantized feedback, and vanishes entirely as $P \rightarrow \infty$. However, for analog feedback, the decrease is relatively slow (i.e.\@ only polynomially fast) and does not vanish as $P \rightarrow \infty$. The analysis may also be extended to the case when errors occur with quantized feedback, using techniques similar to those in \cite{caire2007}.

\subsection{Generation of Numerical Results}
\label{sec:simmethod}

The number of bits given by (\ref{eqn:B}) can be very large and numerical simulation becomes a computationally complex task, as the chordal distance will have to be calculated for each of the $2^B$ matrices in the codebook. However, utilizing the statistics of random codebooks, the quantization procedure can be precisely {\emph{emulated}} without having to do actual quantization. From Lemma \ref{lemma1}, we can repeat the argument by interchanging $\widetilde{\Hm}_k$ and $\Hhm_k$, to yield the following equivalent decomposition:
\begin{equation}
\Hhm_k = \widetilde{\Hm}_k \Xm_k \Ym_k + \Sm_k \Zm_k
\end{equation}
which can be used to generate $\Hhm_k$, given $\widetilde{\Hm}_k$ and a codebook size. $\Xm_k$ is isotropic and independent of the codebook size, as is $\Sm_k$ which (in this decomposition) is isotropically distributed in the left nullspace of $\widetilde{\Hm}_k$. Samples drawn from the distribution of these matrices can thus be generated as samples from the isotropic distribution in their respective spaces.

Moreover, $d^2\left( \widetilde{\Hm}_k, \Hhm_k \right) = \trace\left( \Zm_k^\herm\Zm_k \right)$ is the $1^\text{st}$ order statistic from $2^B$ samples. Here, each sample is drawn from the distribution of the trace of a matrix-variate beta distribution (as described in Appendix \ref{lem1proof}). Thus, a sample drawn from the distribution of $\trace\left( \Zm_k^\herm\Zm_k \right)$ can be generated by the `CDF inversion' method, by computing the CDF for a specific $M$ and $N$. A general expression for the CDF has been computed in closed form in \cite{DAI06}, for the case when $d^2\left( \widetilde{\Hm}_k, \Hhm_k \right) \leq 1$. For moderate to large $B$ and practical values of $M$, $N$, this event occurs with extremely high probability, allowing for low complexity CDF inversion. For very small values of $B$, $d^2\left( \widetilde{\Hm}_k, \Hhm_k \right)$ may be greater than 1 with appreciable probability, but an exhaustive searching among $2^B$ possibilities is not a problem in these cases.

From the eigen decomposition $\Zm_k^\herm\Zm_k = \Em_k\Dm_k\Em_k^\herm$, as described in Appendix \ref{thm1proof}, $\Em_k$ can be generated as the eigenvectors of any (complex) Beta$(N, M-N)$ distributed matrix. Further, the distribution of the eigenvalues (i.e.\@, the entries of $\Dm_k$) {\emph{conditioned}} on their sum (which is equal to $d^2(\widetilde{\Hm}_k, \Hhm_k)$), can be computed from their joint distribution \cite{muirhead1982ams} (\cite{DAI06} for the complex case). The conditional distribution can be easily computed for small values of $N$.

In particular, for $N = 2$, if $D_1, D_2$ are the diagonal elements of $\Dm_k$ with joint density $f_{D_1, D_2}(d_1, d_2)$, the distribution of $D_1$ conditioned on $Z = D_1 + D_2 \leq 1$ is given as:
\begin{eqnarray}
F_{D_1|Z}(d_1|z) & = & \frac{\int\limits_{0}^z f_{D_1, D_2}(d_1, z-d_1)\ {d}(d_1)}{f_Z(z)}\\
& = & \frac{\int\limits_{0}^z V_{M} (z-2d_1)^2(1-d_1)^{M-4}(1-z+d_1)^{M-4}\ {d}(d_1)}{f_Z(z)}
\end{eqnarray}
where $f_Z(z)$ is the pdf of $Z$ computed to be:
\begin{equation}
f_Z(z) = \frac{z^{2M-5} (\Gamma(M))^2}{(M-1)\Gamma(2M-4)}
\end{equation}
for $z \leq 1$. $V_M$ is a normalizing constant and is given by $V_M = \frac{1}{2} (M-1)(M-2)^2(M-3)$. For efficient CDF inversion, $F_{D_1|Z}(d_1|z)$ can be computed in closed form for specific values of $M$.

As $\Ym_k^\herm\Ym_k = {\bf I}_N - \Zm_k\Zm_k^\herm$, $\Ym_k$ can be obtained as well. Putting all this together, one is able to randomly generate a realization of the quantized version of $\widetilde{\Hm}_k$, when random codebooks are used. This prevents the computational complexity from growing with $B$. However, for extremely large $B$, numerical errors may dominate and care must be taken to maintain numerical precision.

\subsection{Numerical Results}
\label{sec:simresults}

\begin{figure}[htb]
\begin{center}
\includegraphics[width=\ImgWidth]{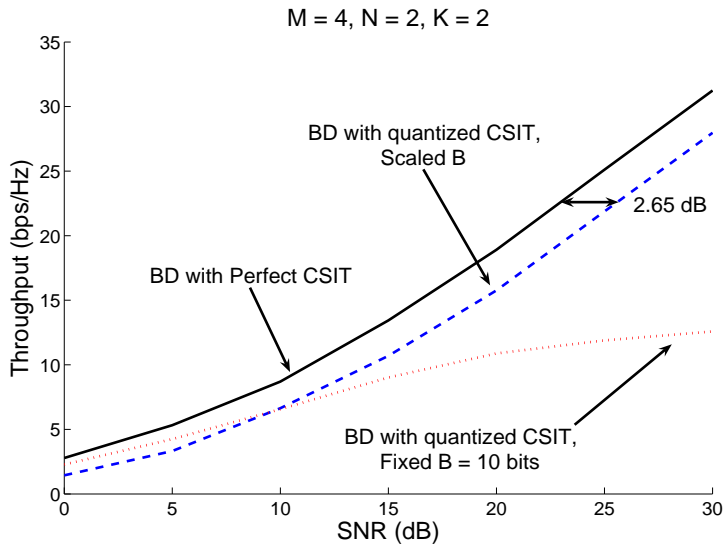}
\caption{MIMO Broadcast Channel with $M = 4, N = 2, K = 4$}\label{fig1}
\end{center}
\end{figure}

\begin{figure}[htb]
\begin{center}
\includegraphics[width=\ImgWidth]{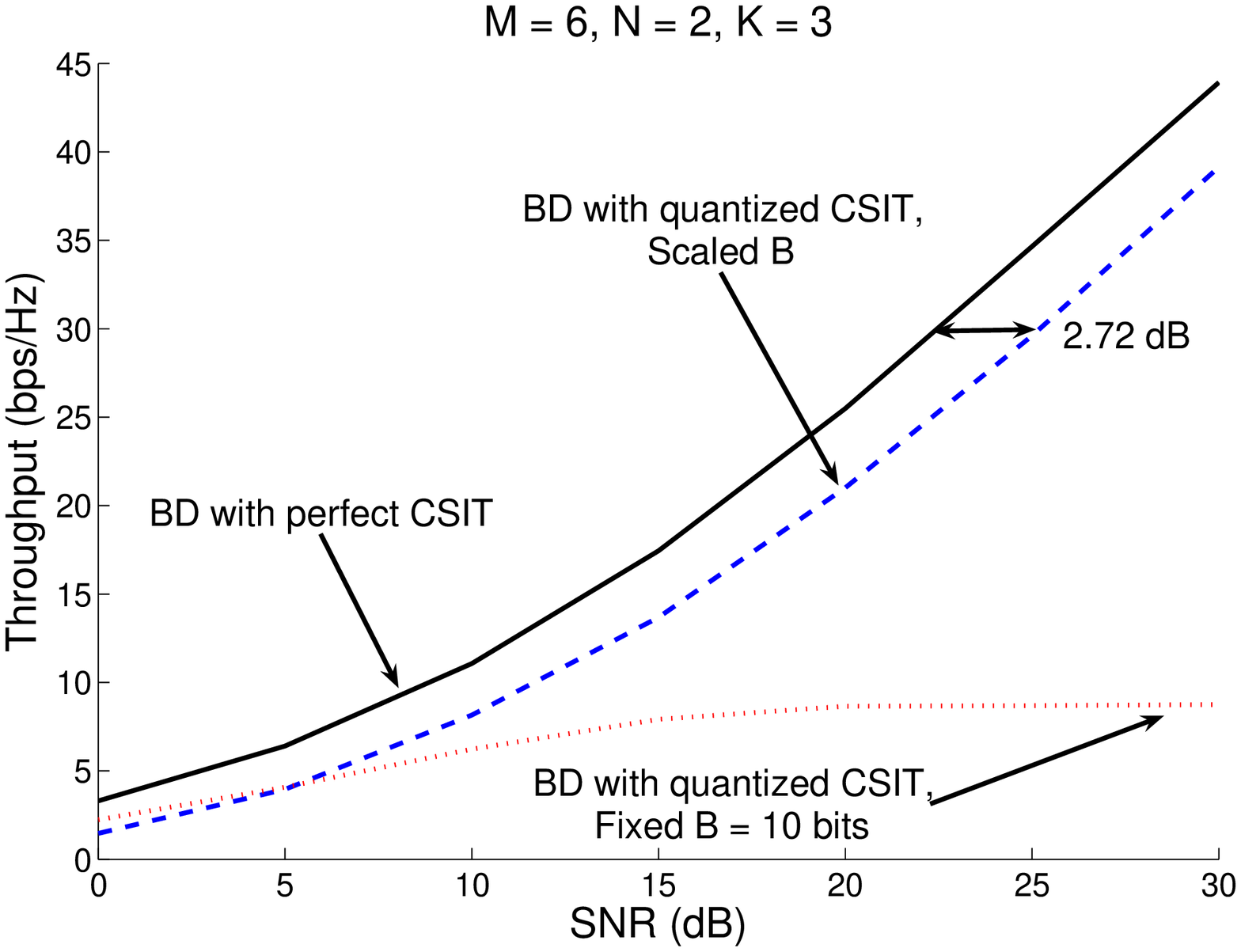}
\caption{MIMO Broadcast Channel with $M = 6, N = 2, K = 4$}\label{fig2}
\end{center}
\end{figure}

\begin{figure}[htb]
\begin{center}
\includegraphics[width=\ImgWidth]{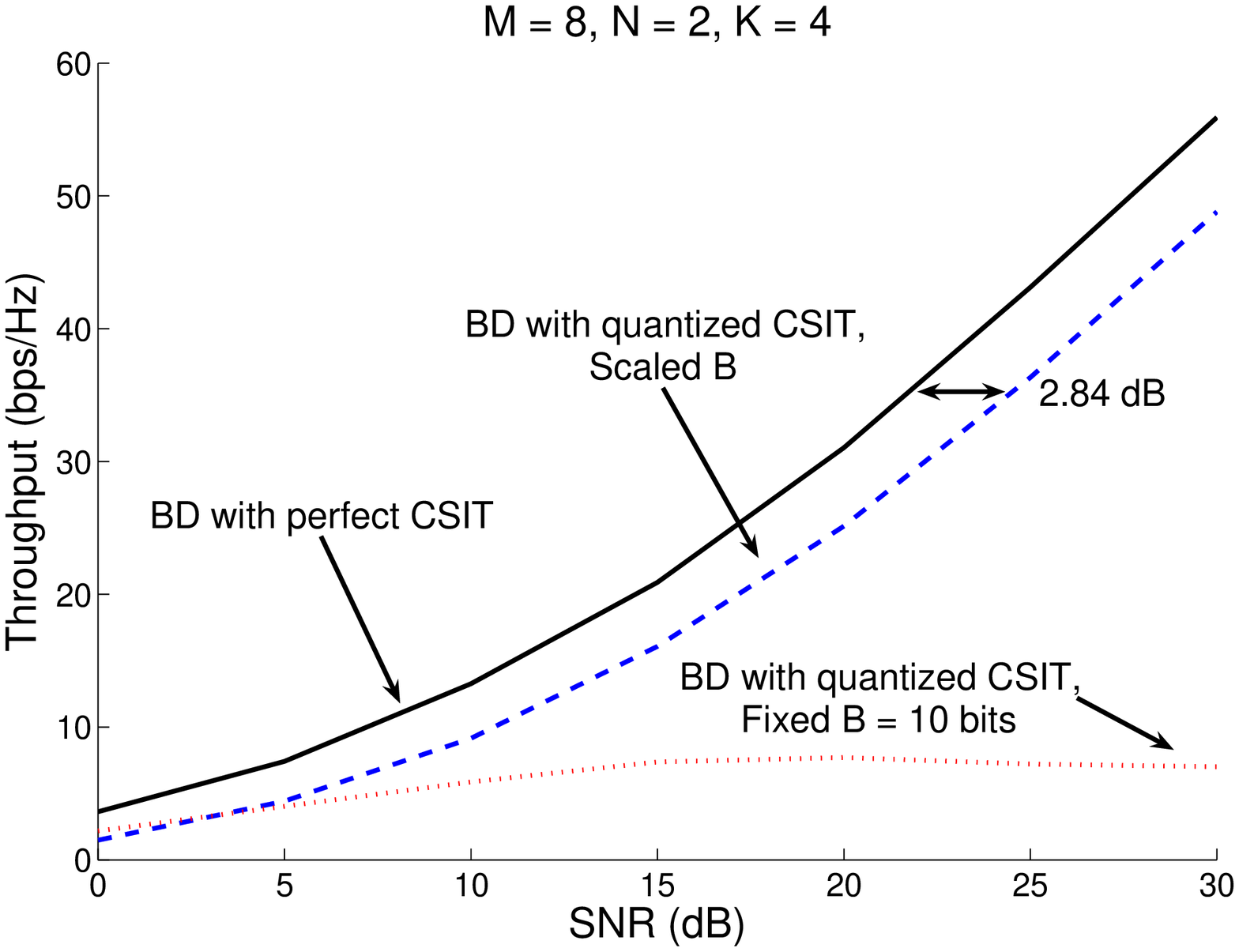}
\caption{MIMO Broadcast Channel with $M = 8, N = 2, K = 4$}\label{fig3}
\end{center}
\end{figure}

We present numerical results for $N=2$ and $M = 4, 6, 8$ in Figures \ref{fig1}, \ref{fig2}, \ref{fig3} respectively, while scaling the bits as per (\ref{eqn:BDSca}), i.e.\@ with a target of staying at most 3 dB away (in SNR) from BD with perfect CSIT. As Theorem \ref{thm:2} only provides the sufficient number of bits, this is a conservative strategy and the actual SNR gaps are found to be $2.65$ dB, $2.72$ dB and $2.84$ dB for $M = 4, 6$ and $8$ respectively, instead of $3$ dB. The results also show that keeping the number of bits fixed will result in a rate gap that increases unbounded with SNR.

\section{Conclusion}
\label{sec:concl}

Accurate CSIT is clearly important for MIMO broadcast systems in order to achieve maximum throughput. When the receiver knows the channel perfectly and instantaneously feeds this information back to the transmitter using a finite number of bits, we have quantified the rate loss and have shown that increasing the number of bits linearly with the system SNR is sufficient to maintain a constant SNR loss with respect to perfect CSIT. Further, we have established the advantage of BD relative to ZF in terms of feedback load, and the advantage of using quantized feedback as opposed to using analog feedback. Note that BD is just one of many linear precoding techniques that can be used on the MIMO broadcast channel with multiple user antennas (for e.g.\@, see coordinated beamforming \cite{chae2006cbm} and Multiuser Eigenmode Transmission \cite{boccardi2007not}). It remains to be seen which of these perform best in a limited feedback setting and also when multiuser diversity/user selection is considered.

\appendices

\section{Proof of Lemma \ref{lemma1}} \label{lem1proof}

Let $\Wm$ be any arbitrary matrix in the codebook $\Cc$. Note that $\Wm$ is independent of $\widetilde{\Hm}_k$. We then decompose $\widetilde{\Hm}_k$ into components that lie in the column space of $\Wm$ and the left nullspace of $\Wm$ as follows:
\begin{eqnarray}
\widetilde{\Hm}_k & = & \Wm\Wm^\herm \widetilde{\Hm}_k + \left( {\bf I}_M - \Wm\Wm^\herm \right) \widetilde{\Hm}_k \\
& = & \Wm\Wm^\herm \widetilde{\Hm}_k + \Wm^\bot(\Wm^\bot)^\herm \widetilde{\Hm}_k
\end{eqnarray}
where $\Wm\Wm^\herm$ and $\Wm^\bot(\Wm^\bot)^\herm = {\bf I}_M - \Wm\Wm^\herm$ are the projection matrices for the column space and left nullspace of $\Wm$ respectively. $\Wm^\bot \in \CC^{M \times (M-N)}$ is chosen such that it forms an orthonormal basis for the left nullsapce of $\Wm$.

Let the (thin) QR decomposition of $\Wm\Wm^\herm \widetilde{\Hm}_k$ be $\Qm_k {\Am}_k$ where $\Qm_k \in \CC^{M \times N}$ forms an orthonormal basis for the same space as $\Wm$, and $\Am_k \in \CC^{N \times N}$ is upper triangular with positive diagonal elements. Further, $\Qm_k$ and $\Am_k$ are independent, from \cite[Theorem 2.3.18]{gupta2000mvd} (after verification for the complex case). As $\Qm_k$ and $\Wm$ describe the same subspace, $\Qm_k$ may be represented as a rotation of $\Wm$, i.e.\@, $\Qm_k = \Wm \Xm_k$ for some unitary matrix $\Xm_k \in \CC^{N \times N}$. 

By isotropy and independence of $\Wm$ and $\widetilde{\Hm}_k$, $\Xm_k$ is also isotropically distributed and is independent of $\Wm$, which is an arbitrary orthonormal basis. Also note that $\Wm\Wm^\herm = \Qm_k\Qm_k^\herm$ and hence $\Am_k^\herm\Am_k = \widetilde{\Hm}_k^\herm\Wm\Wm^\herm\widetilde{\Hm}_k$. Thus $\trace\left(\Am_k^\herm\Am_k\right) = N - d^2\left(\Wm, \widetilde{\Hm}_k\right)$.

Note that $\Wm^\bot(\Wm^\bot)^\herm \widetilde{\Hm}_k$ is the projection of $\widetilde{\Hm}_k$ onto the left nullspace of $\Wm$. As $\widetilde{\Hm}_k$ is isotropically distributed, the projection is also isotropically distributed in the corresponding $M-N$ dimensional nullspace. Let the (thin) QR decomposition of $\Wm^\bot(\Wm^\bot)^\herm \widetilde{\Hm}_k$ be $\Sm_k \Bm_k$, where $\Sm_k \in \CC^{M \times N}$ is an orthonormal basis for an isotropically distributed (complex) $N$ dimensional plane in the $M-N$ dimensional left nullspace of $\Wm$ and $\Bm_k \in \CC^{N \times N}$ is upper triangular with positive diagonal elements. Similar to the previous case, $\Sm_k$ and $\Bm_k$ are independently distributed. It is also straightforward to see that $\Bm_k^\herm\Bm_k = {\bf I}_N - \Am_k^\herm\Am_k$ and $\trace\left(\Bm_k^\herm\Bm_k\right) = d^2\left(\Wm, \widetilde{\Hm}_k\right)$.

As $\widetilde{\Hm}_k$ and $\Wm$ are independent, which has been our assumption thus far in the proof, $\Bm_k^\herm\Bm_k$ is matrix-variate (complex) Beta$(N, M-N)$ distributed \cite{muirhead1982ams}. We will now argue that most of the above conclusions remain unchanged, even when the quantization procedure (\ref{quantproc}) is followed.

The quantization procedure amounts to choosing a $\Bm_k^\herm\Bm_k$ such that its trace is the minimum among $2^B$ choices. Thus, it follows that the quantization procedure only affects $\Bm_k$ (and $\Am_k$, which is the `inverse' quantization error and is related to $\Bm_k$ by $\Am_k^\herm\Am_k = {\bf I}_N - \Bm_k^\herm\Bm_k$). We use $\Ym_k$ and $\Zm_k$ to denote the matrices $\Am_k$ and $\Bm_k$ after following the quantization procedure. Hence, even though $\Zm_k^\herm\Zm_k$ is not beta distributed, the distribution of the quantities $\Xm_k$, $\Sm_k$ and $\Wm$ remain the same, and are independent of $\Zm_k$ (and $\Ym_k$). We now use $\Hhm_k$ to denote $\Wm$ after following the quantization procedure, according to the convention in (\ref{quantproc}).

\section{Proof of Theorem \ref{thm:1}} \label{thm1proof}

Theorem \ref{thm:1} is proved as follows:
\begin{eqnarray}
\Delta R_\textsc{Quant}(P) & = & \left[  R_\textsc{CSIT-BD}(P) - R_\textsc{Quant}(P)  \right]\\
& \mathop{\leq}\limits^{\text{\tiny{(a)}}} & \EE \left[ \log_2 \left| {\bf I}_N + \frac{P}{M}\Hm_k^\herm\Vm_k\Vm_k^\herm\Hm_k \right| \right] - \nonumber\\
& & \EE \left[ \log_2 \left| {\bf I}_N + \frac{P}{M}\ \Hm_k^\herm\Vhm_k\Vhm_k^\herm\Hm_k \right| \right] + \nonumber\\
& & \EE \left[ \log_2 \left| {\bf I}_N + \frac{P}{M} \sum\limits_{j = 1, j \neq k}^K \Hm_k^\herm\Vhm_j\Vhm_j^\herm\Hm_k \right| \right] \\
& \mathop{=}\limits^{\text{\tiny{(b)}}} & \EE \left[ \log_2 \left| {\bf I}_N + \frac{P}{M} \sum\limits_{j = 1, j \neq k}^K\Hm_k^\herm\Vhm_j\Vhm_j^\herm\Hm_k \right| \right]\\
& \mathop{=}\limits^{\text{\tiny{(c)}}} & \EE \left[ \log_2 \left| {\bf I}_N + \frac{P}{M} \widetilde{\Hm}_k^\herm \left( \sum\limits_{j \neq k} \Vhm_j\Vhm_j^\herm  \right) \widetilde{\Hm}_k {\bf \Lambda}_k \right| \right]\\
& \mathop{\leq}\limits^{\text{\tiny{(d)}}} & \log_2\ \left| {\bf I}_N + \frac{P(K-1)}{M} \EE \left[  \widetilde{\Hm}_k^\herm\left( \Vhm_j\Vhm_j^\herm \right)\widetilde{\Hm}_k  \right] M\right|\\
& \mathop{=}\limits^{\text{\tiny{(e)}}} & \log_2\ \left| {\bf I}_N + P(K-1) \EE \left[ \Zm_k^\herm \left( \Sm_k^\herm\Vhm_j \Vhm_j^\herm\Sm_k\right) \Zm_k \right] \right|\\
& \leq & N\ \log_2 \left(1 + \frac{P}{N} D \right)
\end{eqnarray}

Here, (a) follows by neglecting the positive semi-definite interference terms in the quantity:
\begin{equation*}
\EE \left[ \log_2 \left| {\bf I}_N + \frac{P}{M} \sum\limits_{j = 1}^K \Hm_k^\herm\Vhm_j\Vhm_j^\herm\Hm_k \right| \right].
\end{equation*}

\noindent
By the BD procedure, both $\Vm_k$ and $\Vhm_k$ are distributed isotropically, and are chosen independent of $\Hm_k$, which results in (b). We write $\Hm_k\Hm_k^\herm = \widetilde{\Hm}_k {\bf \Lambda}_k \widetilde{\Hm}_k^H$, where $\widetilde{\Hm}_k \in \CC^{M \times N}$ forms an orthonormal basis for the subspace spanned be the columns of $\Hm_k$ and ${\bf \Lambda}_k = \text{diag}[\lambda_1, \dots, \lambda_N]$ are the $N$ non-zero, unordered eigenvalues of $\Hm_k\Hm_k^\herm$ ($\Hm_k$ is of rank $N$ and diagonalizable with probability 1). Both the density function of $\Hm_k$ (which is matrix-variate complex Normal distributed) \cite{gupta2000mvd} and the Jacobian of the singular value decomposition transformation of a matrix \cite{edelman2005rmt} can be separated into a product of functions of $\widetilde{\Hm}_k$ and ${\bf \Lambda}_k$ alone. Thus, $\widetilde{\Hm_k}$ and ${\bf \Lambda}_k$ are independent and $\EE \left[ {\bf \Lambda}_k \right] = M{\bf I}_N$. Step (c) follows using this and the fact that $\left|{\bf I} + \Am\Bm \right|$ = $\left|{\bf I} + \Bm\Am \right|$, for matrices $\Am$ and $\Bm$. Next, (d) follows from Jensen's inequality due to the concavity of $\log|\cdot|$. Step (e) is proved as follows. First, we compute
\begin{eqnarray}
\widetilde{\Hm}_k^\herm \Vhm_j & = & \Ym_k^\herm\Xm_k^\herm\Hhm_k^\herm\Vhm_j + \Zm_k^\herm\Sm_k^\herm\Vhm_j\\
& = &  \Zm_k^\herm\Sm_k^\herm\Vhm_j
\end{eqnarray}
for $k \neq j$, which follows from Lemma \ref{lemma1} and the fact that $\Hhm_k^\herm\Vhm_j = {\bf 0}\ \forall k \neq j$, by the BD procedure.
Therefore,
\begin{eqnarray}
\log_2\ \left| {\bf I}_N + P(K-1) \EE \left[ \widetilde{\Hm}_k^\herm \Vhm_j\Vhm_j^\herm \widetilde{\Hm}_k \right] \right| & = & \log_2\ \left| {\bf I}_N + P(K-1) \EE \left[ \Zm_k^\herm \left( \Sm_k^\herm \Vhm_j\Vhm_j^\herm \Sm_k\right) \Zm_k \right] \right|\nonumber\\
& \mathop{=}\limits^{\text{\tiny{(f)}}} & \log_2\ \left| {\bf I}_N + P(K-1) \frac{N}{M-N} \EE \left[ \Zm_k^\herm\Zm_k \right] \right|\\
& \mathop{=}\limits^{\text{\tiny{(g)}}} & \log_2\ \left| {\bf I}_N + P(K-1) \frac{D}{N} \frac{N}{M-N} \right|
\end{eqnarray}
Here, (f) follows from the fact that $\Vhm_j$ (which is just isotropically distributed in the left nullspace of $\Hhm_k$) and $\Zm_k$ are independent, as are $\Sm_k$ and $\Zm_k$ from Lemma \ref{lemma1}. Further, $\Sm_k$ is also isotropically and distributed in the left nullspace of $\Hhm_k$, and is independent of $\Vhm_k$. Thus $\Vhm_j^\herm \Sm_k \Sm_k^\herm \Vhm_j$ is matrix-variate Beta$(N, M-2N)$ distributed \cite{gupta2000mvd}, and $\EE \left[ \Zm_k^\herm \left( \Sm_k^\herm \Vhm_j\Vhm_j^\herm \Sm_k\right) \Zm_k \right] = \frac{N}{M-N} \EE\left[ \Zm_k^\herm\Zm_k \right]$, by \cite[Theorem 5.3.12]{gupta2000mvd} and \cite[Theorem 5.3.19]{gupta2000mvd} (after verification for the complex case).

Let $\Em_k\Dm_k\Em_k^\herm$ be the eigen decomposition of $\Zm_k^\herm\Zm_k$, where $\Em_k \in \CC^{N \times N}$ is orthonormal and $\Dm_k \in \CC^{N \times N}$ is diagonal, with strictly positive elements along the diagonal. If an arbitrary matrix in the codebook $\Cc$ is selected as the quantization, $\Zm_k^\herm\Zm_k$ is matrix-variate (complex) Beta$(N, M-N)$ distributed (as described in Appendix \ref{lem1proof}), and $\EE\left[\Zm_k^\herm\Zm_k \right]$ is a multiple of the identity matrix. Both the density function of this distribution \cite{gupta2000mvd} and the Jacobian of the eigen decomposition transformation for a matrix \cite{edelman2005rmt} can be separated into a product of functions of $\Em_k$ and $\Dm_k$ alone, and these are hence independently distributed. 

For the actual quantization matrix, after following the procedure in (\ref{quantproc}), only the distribution of the diagonal matrix $\Dm_k$ is affected, and the distribution of $\Em_k$ remains unchanged and independent of $\Dm_k$. Thus, we have that $\EE\left[ \Zm_k^\herm\Zm_k \right] = \rho {\bf I}_N$ for some constant $\rho$, even after following the quantization procedure. This can also be concluded by observing that $\Zm_k^\herm\Zm_k$ is invariant to unitary rotations. In terms of the trace of the matrix, we have $\rho = \frac{\EE\left[\trace\left(\Zm_k\Zm_k^\herm\right)\right]}{N} = \frac{D}{N}$, and (g) follows.

\section{Proof of equation (\ref{analogbound})} \label{analogproof}

$\Delta R_\textsc{Analog}(P) = \left[  R_\textsc{CSIT-BD}(P) - R_\textsc{Analog}(P)  \right]$
\begin{eqnarray}
& \mathop{\leq}\limits^{\text{\tiny{(a)}}} & \EE \left[ \log_2 \left| {\bf I}_N + \frac{P}{M}\Hm_k^\herm\Vm_k\Vm_k^\herm\Hm_k \right| \right] - \nonumber\\
& & \EE \left[ \log_2 \left| {\bf I}_N + \frac{P}{M}\ \Hm_k^\herm\breve{\Vm}_k\breve{\Vm}_k^\herm\Hm_k \right| \right] + \nonumber\\
& & \EE \left[ \log_2 \left| {\bf I}_N + \frac{P}{M} \sum\limits_{j = 1, j \neq k}^K \Hm_k^\herm\breve{\Vm}_j\breve{\Vm}_j^\herm\Hm_k \right| \right] \\
& \mathop{=}\limits^{\text{\tiny{(b)}}} & \EE \left[ \log_2 \left| {\bf I}_N + \frac{P}{M} \sum\limits_{j = 1, j \neq k}^K\Hm_k^\herm\breve{\Vm}_j\breve{\Vm}_j^\herm\Hm_k \right| \right]\\
& \mathop{=}\limits^{\text{\tiny{(c)}}} & \EE \left[ \log_2 \left| {\bf I}_N + \frac{P}{M} \frac{1}{1 + \beta P} \sum\limits_{j = 1, j \neq k}^K\Fm_k^\herm\breve{\Vm}_j\breve{\Vm}_j^\herm\Fm_k \right| \right]\\
& \mathop{\leq}\limits^{\text{\tiny{(d)}}} & \log_2\ \left| {\bf I}_N + \frac{P(K-1)}{M}  \frac{1}{1 + \beta P} \EE \left[  \Fm_k^\herm\breve{\Vm}_j\breve{\Vm}_j^\herm\Fm_k  \right] \right|\\
& \mathop{=}\limits^{\text{\tiny{(e)}}} & \log_2\ \left| {\bf I}_N + \frac{P(K-1)}{M}  \frac{1}{1 + \beta P} N {\bf I}_N \right|\\
& = & N\ \log_2 \left(1 + \frac{M-N}{M}  \frac{P}{1 + \beta P} \right)
\end{eqnarray}
Here, (a) and (b) have the same justification as in the proof of Theorem \ref{thm:1} (in Appendix \ref{thm1proof}), (c) follows from (\ref{int_ana}), and (d) is obtained by applying Jensens inequality. By Gaussianity of $\Fm_k$ and independence of $\Fm_k$ and $\breve{\Vm}_j$, $\Fm_k^\herm\breve{\Vm}_j$ is matrix-variate complex Gaussian distributed with i.i.d.\@ elements, and $\EE \left[  \Fm_k^\herm\breve{\Vm}_j\breve{\Vm}_j^\herm\Fm_k \right] = N {\bf I}_N$, which results in (e).

\newpage
\bibliographystyle{IEEETran}
\bibliography{BD_JSAC}

\end{document}